\documentclass[3p, twocolumn]{elsarticle}

\usepackage{color}
\usepackage{url}
\usepackage{amsmath}
\usepackage{comment}
\usepackage{graphicx}

\begin{document}

\pagestyle{headings}

\begin{frontmatter}
\title{Detection of money laundering groups using supervised learning in networks}

\author[RMITCSIT]{David Savage\corref{cauthor}}
\cortext[cauthor]{Corresponding author}
\ead{david.savage@rmit.edu.au}

\author[RMITECE]{Qingmai Wang}

\author[AUSTRAC]{Pauline Chou}
\author[RMITCSIT]{Xiuzhen Zhang}
\author[RMITECE]{Xinghuo Yu}

\address[RMITCSIT]{School of Computer Science and Information Technology, RMIT University, Melbourne, Australia}
\address[RMITECE]{School of Electrical and Computer Engineering, RMIT University, Melbourne, Australia}
\address[AUSTRAC]{Australian Transaction Reports and Analysis Centre, PO Box 13173, Law Courts, Melbourne, Victoria 8010, Australia}

%


\begin{abstract}

Money laundering is a major global problem, enabling criminal organisations to hide their ill-gotten gains and to finance further operations. Prevention of money laundering is seen as a high priority by many governments, however detection of money laundering without prior knowledge of predicate crimes remains a significant challenge. Previous detection systems have tended to focus on individuals, considering transaction histories and applying anomaly detection to identify suspicious behaviour. However, money laundering involves groups of collaborating individuals, and evidence of money laundering may only be apparent when the collective behaviour of these groups is considered. In this paper we describe a detection system that is capable of analysing group behaviour, using a combination of network analysis and supervised learning. This system is designed for real-world application and operates on networks consisting of millions of interacting parties. Evaluation of the system using real-world data indicates that suspicious activity is successfully detected. Importantly, the system exhibits a low rate of false positives, and is therefore suitable for use in a live intelligence environment.

\end{abstract}


\end{frontmatter}

\section{Introduction}

This paper describes an automated system for detecting money laundering groups in a financial transaction network. This system employs a combination of network analysis and supervised learning to identify suspicious behaviour indicative of money laundering activity. The system is designed for use in a live intelligence environment at the Australian Transaction Reports and Analysis Centre (AUSTRAC).

Money laundering is a major global problem, with numerous detrimental impacts on society. Most importantly, money laundering enables organised criminal groups to flourish, leading to increased incidents of predicate crime (i.e. crimes that generate the funds to be laundered). Moreover, significant levels of money laundering can severely undermine economies and financial systems \cite{Levi:2006ir,AUSTRAC:2011ve}.

The process of money laundering involves three main stages termed placement, layering, and integration \cite{AUSTRAC:2011ve}. The placement stage represents the introduction of funds obtained through criminal activities into the financial system. Typically, this involves deposits that are spread over time and geographical locations. Once the funds have been placed into the financial system, layering is undertaken in order to hide the original source of the funds. This stage include numerous transactions, and often involves offshore accounts and complex investment vehicles. In the final stage, integration, funds (or equivalent value) are transferred to the owners, often in the form of investments or tangible goods (e.g. jewellery, high-end cars, etc.).

It is clear from the three stages of money laundering that a typical operation will involve multiple transactions, conducted through a variety of different channels, by a group of parties (individuals, businesses, etc.) acting in collusion \cite{Irwin:2011it,He:2010it,AUSTRAC:2011ug}. However, existing systems tend to focus on individual parties, considering individual transaction histories and applying anomaly detection to identify suspicious behaviour (e.g. \cite{Zhu:2006uc,LeKhac:2010kz,Zengan:2009uf,Raza:2011cq,Kingdon:2004ii}). In this paper, we describe a system that advances the current state-of-the-art in this area, considering the collective behaviour of small groups represented as communities in a transaction network.



\subsection{Summary of Contribution}

The system described in this paper represents an end-to-end solution for automated detection of money laundering activity. The system is designed to run as an ongoing monitoring tool in a live environment and is expected to analyse millions of transactions. This includes the construction of a network model representing relationships derived from financial records held by AUSTRAC, extraction of meaningful communities from this network, generation of features capturing the key characteristics of these communities, and finally, classification using a supervised learning approach. The major novelty of this system stems from two main considerations.

\begin{itemize}
\item{\textbf{Network analysis combining financial transactions and supplementary relationships.} The network analysed by our system contains multiple relationships, represented using typed edges.  In addition to the actual remittance of funds, parties may be linked by shared accounts, shared use of agents, overlapping geolocations, etc. In determining the strength of a connection between two parties, different types of relationships are weighted to reflect perceived importance. This allows business knowledge to be incorporated into the network model.}
\item{\textbf{Treatment of near-k-step neighbourhoods as observations for supervised classifiers.} Comparable systems described in the literature have focused on individual parties, typically analysing each party’s transaction history in isolation. Our system considers groups of transacting parties as the basic unit of analysis, extending the notion of `know your customer' to a network setting. As we describe in Section \ref{SecCommunityExtraction}, groups of tightly interacting parties may be extracted from a network in different ways. We have elected to use a bottom up approach for this task, extracting relevant parties from a small region centred on each new transaction. This approach is particularly suited to the operational needs of a near-real-time intelligence environment.}
\end{itemize}

In the following sections we describe previous work in this area, provide an overview of our system and evaluate its performance. Results of this evaluation demonstrate the ability of our system to identify suspicious behaviour in real data.

\section{Related Work}

One of the earliest systems for detection of money laundering is that described by Senator \cite{Senator:1995dv}, which applied rule-based evaluation to identify suspicious parties. The rules used by this system were derived from expert knowledge and encoded in an evaluation module that was run each time the target database was updated. Parties matching these rules would then be further investigated by analysts using an interactive query interface and a variety of visualisation tools provided by the system. More recently, Wang et al. describe an alternative rules-based system, where rules are encoded using a decision tree \cite{Wang:2007dd}.

While rule-based systems may be highly accurate, they are dependent on expert knowledge, and cannot be used to uncover new typologies (i.e. modes of operation). Later systems address this issue by applying more flexible approach based on a combination of supervised and unsupervised learning. Many of these systems follow a basic premise, first set out by Kingdon \cite{Kingdon:2004ii}, which centres on the notion of ‘know your customer’ and the use of anomaly detection for identifying money laundering behaviour.

In these systems, two main contexts are considered for deriving models of normal, non-suspicious behaviour. The first context is provided by the transaction history for a given party, while the second context is provided by sets of parties exhibiting similar behaviour. In the original system described by Kingdon, grouping of parties into related sets was based on a small number of superficial features such as the use of similar banking products, or sets of businesses providing the same service. Later systems have greatly improved on this scheme, applying distance-based clustering across a far broader range of features (e.g. \cite{Zhu:2006uc,LeKhac:2010kz,Zengan:2009uf,Raza:2011cq}).

In contrast to an anomaly detection approach, a number of systems apply supervised learning to identify suspicious behaviour \cite{Lv:2008tw,Heidarinia:2014te}. In general, these systems are expected to provide a higher degree of precision than those based on anomaly detection, since anomalous behaviour does not necessarily translate to money laundering activity. However, unlike those systems based on anomaly detection, these systems can only identify suspicious behaviour that is similar to that observed in previous investigations.

To date, the majority of systems reported in the literature have focused on individual parties, considering amounts transacted, frequency of transactions, etc. However, more recent studies have begun to adopt a network-based approach, considering features derived from the structure of a transaction network.

For example, the system described in \cite{Drezewski:2015fx} uses role assignment to augment a more traditional approach based on anomaly detection (described in \cite{Drezewski:2012jt}). Using bank statements, a social network is constructed with parties linked by transactions. For each party in the network a number of invariants (betweenness centrality, page rank, etc.) are calculated. A role is then assigned to the party depending on the values of these invariants. Examples of roles include \textit{insulators}, who act as a buffer between a core group of parties and the larger network, and \textit{communicators} who act as a conduit for movement of funds between two otherwise unconnected parties. Assigned roles are then taken into account when considering the normality of a given parties transactions, with parties having the same role expected to show similarities in their transaction histories.

Taking the structural considerations even further, the systems described in \cite{Bershtein:2013hn,Michalak:2011vs} aim to identify subgraphs within a network that closely match known typologies . In these systems, the use of fuzzy matching means that subgraphs may deviate in some way from the given typology, providing greater flexibility than a simple motif search.

\section{System Description}

\begin{figure*}
\centering
\includegraphics[keepaspectratio, width=\textwidth,trim={2cm 21.5cm 2cm 1.5cm},clip]{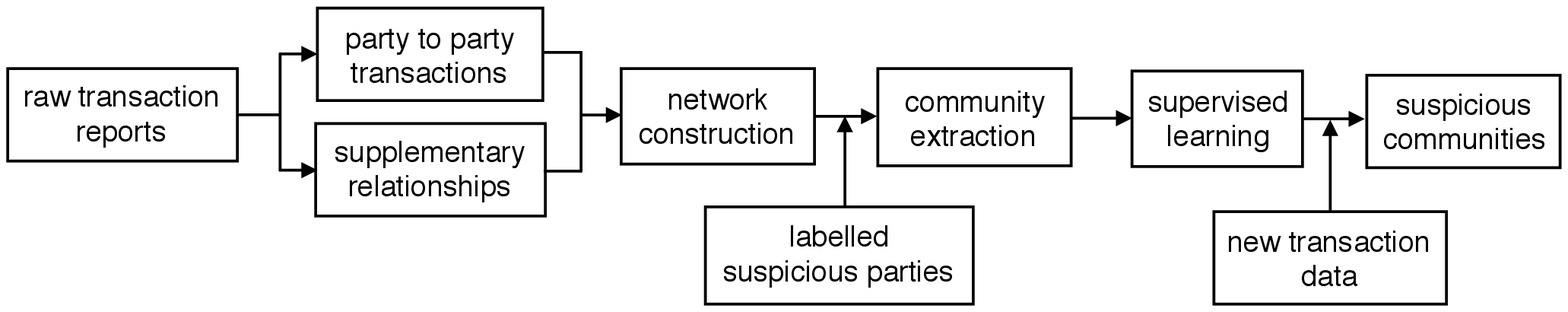}
\caption{System overview.}
\label{FigOverview}
\end{figure*}

Our system performs four main tasks. (1) Modelling of relationships derived from AUSTRAC data as a typed, attributed network. (2) Extraction of communities from the transaction network. (3) Calculation of features from extracted communities, capturing information related to transaction dynamics, party demographics and community structure. (4) Supervised machine learning, treating extracted communities as observations. Figure \ref{FigOverview} shows a high level overview of the system.

The system is designed to run on an ongoing basis, analysing new activity as it occurs. Initially, we extract a random set of communities from the transaction network, and combine these with a set of known suspicious communities. This forms the training set for our supervised learning. Having obtained a trained classifier, the system is then employed for analysis of new activity. For each new transaction reported, the initiating party is treated as a seed and the community containing this party is extracted from the network. Selected features are then calculated, and the community is classified as suspicious or non-suspicious using the previously trained classifier. Those communities that are deemed to be suspicious are then passed to intelligence analysts for further investigation.

\subsection{Transaction Network}
\label{SecNetworkDescription}

\begin{figure*}
\centering
\includegraphics[keepaspectratio, width=\textwidth, trim={2cm 20cm 5cm 1cm}, clip]{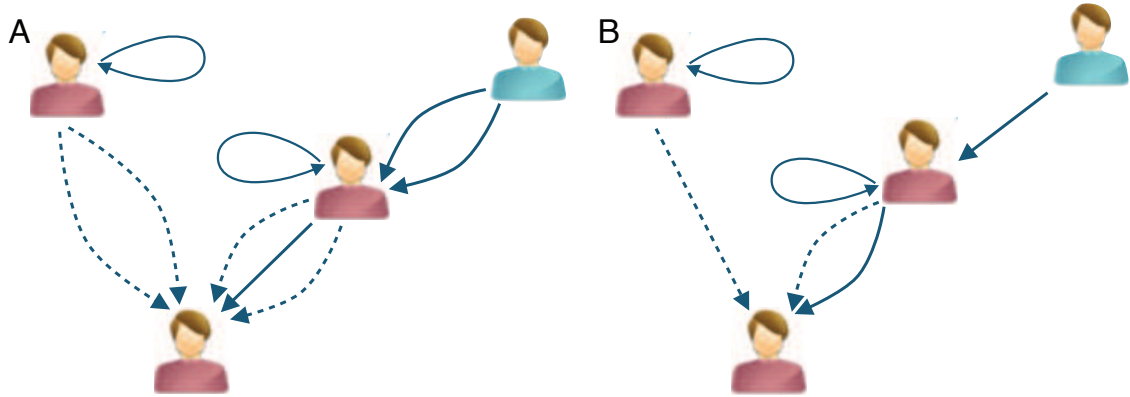}
\caption{Example network consisting of four parties linked by transactions and supplementary evidence. Network A shows the individual transactions (solid edges) and supplementary relationships (dashed edges) connecting the four parties. Network B shows the same four parties with multiple transactions and multiple supplementary relationships summarised as a single edge of each type. Party colours indicate different countries.}
\label{FigExampleNetwork}
\end{figure*}

Our system considers two types of transactions reported to AUSTRAC; large cash deposits and international funds transfers. Reports provide details of each transaction (amounts, requested currency, etc.) and also provide additional information on the sending or receiving parties. Information held in these reports is modelled as a transaction network, with nodes representing transacting parties and edges representing relationships between these parties. Both parties and transaction edges have a number of associated attributes, including the name and address of parties involved, total amount transacted, etc. Relationships are divided into two different types; transactions, representing direct transfer of funds, and supplementary connections, representing mutual association of parties with supplementary evidence (e.g. parties who access the same account are connected). An example network is shown in Figure \ref{FigExampleNetwork}.

Each transaction is modelled as a set of edges connecting sending parties with receiving parties. Certain transactions may include multiple senders and receivers, thus each transaction may be represented by multiple edges connecting all sending parties to all receiving parties. Note that cyclic edges are allowed, as parties may deposit cash into their own account, or transfer funds between accounts held in different countries.

In addition to transactions, parties may be connected through relationships derived from supplementary data available in some reports.  These relationships are represented as a weighted edge between each pair of parties associated through supplementary evidence. For example, if multiple parties access the same account, each party will be connected to every other party accessing the account (forming a clique). For these relationships, each edge represents a summary of all evidence connecting the parties involved. Thus each pair of associated parties is connected by a single supplementary edge. Weights on these edges are calculated as follows.

For a given pair of parties $(p, q)$, the strength of their connection through evidence $e$ is given as
\begin{equation*}
w_{p, q, e} = \frac{n(p, e)}{d_e} \cdot \frac{n(q, e)}{d_e - n(p, e)}\\
\end{equation*}

\noindent where $n(p, e)$ gives the number of transactions where party $p$ is associated with evidence $e$ and $d_e$ gives the total number of transactions involving $e$. A weight $w_{p, q, e}$ can thus be interpreted as the probability of first selecting a random party $p$ that is associated with $e$ based on the number of transactions linking $p$ to $e$, and then choosing a second party $q$ in the same way from the remaining set of parties associated with $e$. The total strength of the connection between $p$ and $q$ is then taken as the maximum weight over all evidence connecting them.\\





For the purposes of this paper, we constructed a sample network consisting of relationships extracted from reports submitted AUSTRAC in 2012. Table \ref{TableNetworkSummary} provides a set of summary statistics for this network. One important characteristic is the high number of discrete connected components. This reflects the fact that the transactions considered only include international funds transfers and large cash deposits. Domestic transactions that do not involve large amounts of cash are not represented in the network. Consequently, the network does not exhibit the same degree of connectedness that is typically observed in social networks (see for example \cite{ugander2011anatomy}). 

Within the sample network, a number of individuals are tagged as being suspicious. These individuals have previously been suspected of involvement in money laundering operations. For the purposes of this paper, communities containing these tagged individuals are labelled as positive observations, as described in Section \ref{SectionMachineLearning}.

\begin{table*}
\centering
\caption{Summary statistics for sample transaction network.}
\begin{tabular}{l l}
	Network Statistics &\\
	\hline
	vertex type & \textit{parties}\\
	edge types & \textit{transactions, associations}\\
	parties & $20,854,744$\\
	total edges & $39,283,144$\\
	supplementary edges & $7,674,102$\\
	transaction edges & $31,609,042$\\
	connected components & $4,654,162$\\
	parties in largest component & $11,650,339$\\
	\hline
\end{tabular}
\label{TableNetworkSummary}
\end{table*}


\subsection{Community Extraction}
\label{SecCommunityExtraction}

Evidence of money laundering involves multiple transactions between different parties. For this reason, we are interested in identifying small sets of interacting parties whose collective behaviour is suspicious. Since our system employs supervised learning, small groups of interacting parties must be extracted from the larger network and treated as observations. There are two main-options for this, community detection (top-down) or an ego-centric approach (bottom-up) based on $k$-step neighbourhoods. For our particular purposes, existing community detection algorithms (see \cite{Fortunato:2010vi} for a comprehensive review) suffer from a number of issues, leaving the ego-centric approach as the favoured option.

One limitation of existing community detection algorithms is their inability to handle heterogeneous networks. The vast majority of these algorithms are designed with a single edge type and single vertex type in mind. Moreover, it is difficult to inject business knowledge into existing approaches, as any attributes of the edges and vertices are typically ignored (however see \cite{Yang:2013ws}). While both of these issues can be addressed to some extent through weighting of edges, it can be difficult to combine information held in numerous attributes and numerous edge types into a single meaningful weight.

Another drawback of existing methods is that they often result in excessively large communities \cite{Leskovec:2008vo,Lancichinetti:2010kj}. In general, meaningful communities are thought to contain less than 150 individuals \cite{Leskovec:2008vo,Allen:2004}, and published typologies indicate that investigation of money laundering operations often focuses on a relatively small number of key parties \cite{AUSTRAC:2011ug}. Using traditional community detection algorithms, smaller communities tend to be found only at the extremities of a network \cite{Leskovec:2008vo,Lancichinetti:2010kj}. Typically, these communities consist of entities that have only recently been added to the network. However, within the core of the network, where the vast majority of interactions take place, many of the detected communities are exceedingly large. For this reason, we take an ego-centric approach, building communities as a bottom-up process.

Given the limitations of community detection, we define communities in our system as near-$k$-step neighbourhoods. A $k$-step neighbourhood is obtained by selecting a subject party $p_0$ (also referred to as the seed) and all parties having a distance of $k$ or less from this seed (i.e. there is a path from the subject $p_0$ to the candidate party $p$ containing $k$ or less edges). The subgraph induced by these parties is the $k$-step neighbourhood. We say near-$k$-step, as two additional constraints are applied beyond that imposed by $k$. These additional constraints further limit the parties and relationships included in the neighbourhood.

\begin{figure*}[t]
\centering
\includegraphics[keepaspectratio, width=\textwidth,trim={1cm 21cm 2cm 2cm},clip]{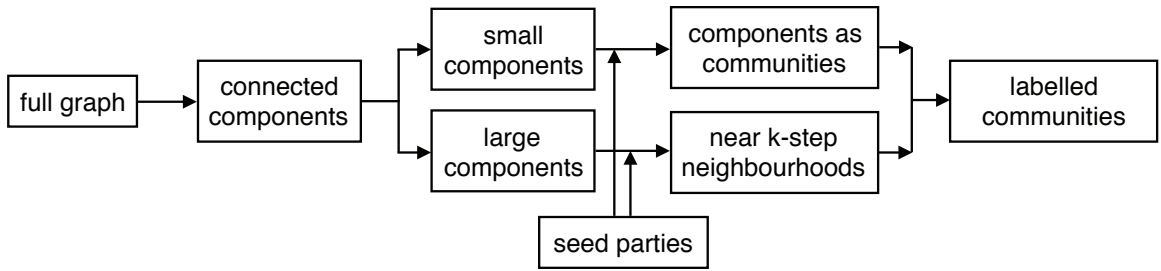}
\caption{Overview of community extraction process.}
\label{FigCommunityOverview}
\end{figure*}

The first constraint is controlled by a parameter, $N_{max}$, which gives the maximum number of transaction neighbours for a party (i.e. the maximum number of parties that have conducted at least one transaction with the subject). Parties having a number of neighbours that exceeds this threshold are treated as gates. Unless linked to other parties within the $k$-step neighbourhood, the transaction neighbours of these parties are not included in the community. This allows us to handle situations, for example, where many parties have transacted with a large corporation but are otherwise unrelated. The second parameter, $w_{min}$, gives the minimum weight for supplementary edges to be included in the community. Note that each of these parameters may be specified as a vector of dimension $k$, so that constraints may vary with distance from the subject party. Figure \ref{FigCommunityOverview} provides an overview of the community detection process.

For the purposes of this paper, we consider $k = 3$, $N_{max} = 40$, and $w_{min} = 0.01$. These values are representative of the actual values used in the live environment, which are selected in consultation with domain experts. In particular our reasoning for taking $k = 3$ stems from the fact that vast majority of transactions in the network are international transfers. By considering 3 steps, we obtain a seed party in the source country, associated parties in the destination country, additional parties in the source country transacting with these same associates, and finally, related parties in the destination country. In other words, setting $k = 3$ allows us to obtain parties in both the source and destination countries that are not directly associated, but are linked through a third party.

As shown in Figure \ref{FigCommunityOverview}, extraction of communities is achieved by first splitting the network into discrete connected components. Those components that are below a threshold size ($diameter \le k$) are treated as communities, with no need to extract the near-$k$-step neighbourhoods. For those components having size greater than the specified threshold, near-$k$-step neighbourhoods are extracted for any subject parties within the component (selection of subject parties is described in Section \ref{SectionMachineLearning}). In our system, this process is implemented using a parallel architecture, allowing millions of communities to be extracted in short period of time.

Treatment of communities as near-$k$-step neighbourhoods can introduce a significant degree of overlap between communities. From an intelligence standpoint, this is a useful characteristic of our approach, as it means the classifier is exposed to numerous views of the same underlying signal, placed within a different context. When assessing a new community, training in this way will mean that the classifier is able to correctly identify suspicious activity even if it is only seeing a small portion of the relevant transactions. One drawback however, is that multiple overlapping communities may be classified as suspicious, and steps must be taken so that analysts are not presented with large numbers of highly similar networks.

For example, suppose that two transactions occur within a short time frame, and that these transactions involve parties whose distance from one another in the network is less than three. For each transaction, a community is extracted using the sending party as the seed. However, since the parties involved are less than three steps from each other, the resulting communities will overlap. If both of these communities are classified as suspicious then without intervention they would both be passed to an analyst (or to two different analysts) for further investigation. Clearly this situation is undesirable, thus a post-processing step is employed that evaluates the degree of overlap between communities deemed to be suspicious. If the overlap is above a certain threshold, then the union of the communities is taken, and it is this result that will be passed to the analyst.

\subsection{Features}

Our system considers a broad range of features derived from the extracted communities. Design of this feature set was guided by expert knowledge (AUSTRAC, unpublished data), and a broad survey of literature related to detection of criminal behaviour and to general network analysis. This feature set is designed to represent different aspects of the transaction network, with features divided into four main categories, as shown in Table \ref{TableFeatureCategories}.


Calculation of dynamic features includes the use of burst analysis, which provides an indication of transaction regularity and is used to identify abnormal behaviour. A wavelet-based algorithm was used for this, which has previously been shown to outperform alternative algorithms in detecting both local and global bursts (Wang, in press).


\begin{table*}
\centering
\caption{Feature categories.}
\begin{tabular}{p{2.5cm} p{9cm}}
category & description\\
\hline
Demographic & Aggregate features describing parties in the network (e.g. mean age)\\
Network & Invariants describing the network structure (e.g. transitivity)\\
Transaction & Aggregations over transactions included in the community (e.g. total cash amount)\\
Dynamic & Features derived from time-series analysis (e.g. num. unusually high amounts)\\
\hline
\end{tabular}
\label{TableFeatureCategories}
\end{table*}

\subsection{Supervised Learning}
\label{SectionMachineLearning}

Our system is designed to allow a high-degree of flexibility in the use of different machine learning models. Given the adversarial nature of the problem domain, and the resulting potential for concept drift, it is important that classifiers are regularly re-evaluated. As notions of suspicious and non-suspicious change over time, new classifiers may need to be trained. In general, this training will include model selection and parameter optimisation. However, for the purposes of this paper, we limit consideration to a support vector machine (SVM) and a random forest, as implemented in the R libraries e1071 and randomForest, respectively. For the SVM we considered a linear and polynomial kernel, using default values for the respective parameters. For the random forest we set the number of trees to one hundred and used default values for all other parameters.

Since the total number of communities that could be extracted from the transaction network is extremely large, training and evaluation was undertaken using a sample of the full set. This sampling, and the assignment of labels, was undertaken as follows. For the labelled true positives we took all parties marked as suspicious in the available database. For each of these parties, we then identified their neighbours in the transaction network and combined these to form a set of positive subjects. For the labelled true negatives, we took a random sample of $700,000$ subjects from those parties not used as subjects for true positives. Near-$k$-neighbourhoods were then extracted for each of the positive and negative subjects. Resulting communities were then assigned the same label as their subject. Note that in certain situations, this process can result in duplicate communities, thus a post-processing step was employed to remove these duplicates.

To obtain a robust estimate of classifier performance, we employed a method similar to $k$-fold cross validation. In each of the $k$ evaluations we took the full set of positive labelled communities, and combined this with the same number of communities randomly sampled from the full set of negative communities. For each of the $k$ iterations, the resulting set was randomly partitioned into training and evaluation sets, with ~70\% of the observations used for training and ~30\% used for evaluation. In each iteration the F-score was calculated using three values for the weighting factor, $\beta_1 = 0.1$, $\beta_2 = 0.5$, $\beta_3 = 1$. These values were selected as we are interested in the ability of the classifier to achieve a high-precision. An ROC curve was also calculated in each iteration and a parameter $\tau$ was determined from this curve giving the threshold value for classifier scores that maximised the F-score for a given value of $\beta$. Precision and recall were then calculated for this value of $\tau$. Mean performance in F-score, precision, recall  and area under the ROC curve was then calculated across the $k$ iterations.


\section{Evaluation}




%

\subsection{Community Extraction}

\begin{figure*}[t]
\centering
\includegraphics[keepaspectratio,width=\textwidth,trim={0cm 0.5cm 2cm 2.2cm},clip]{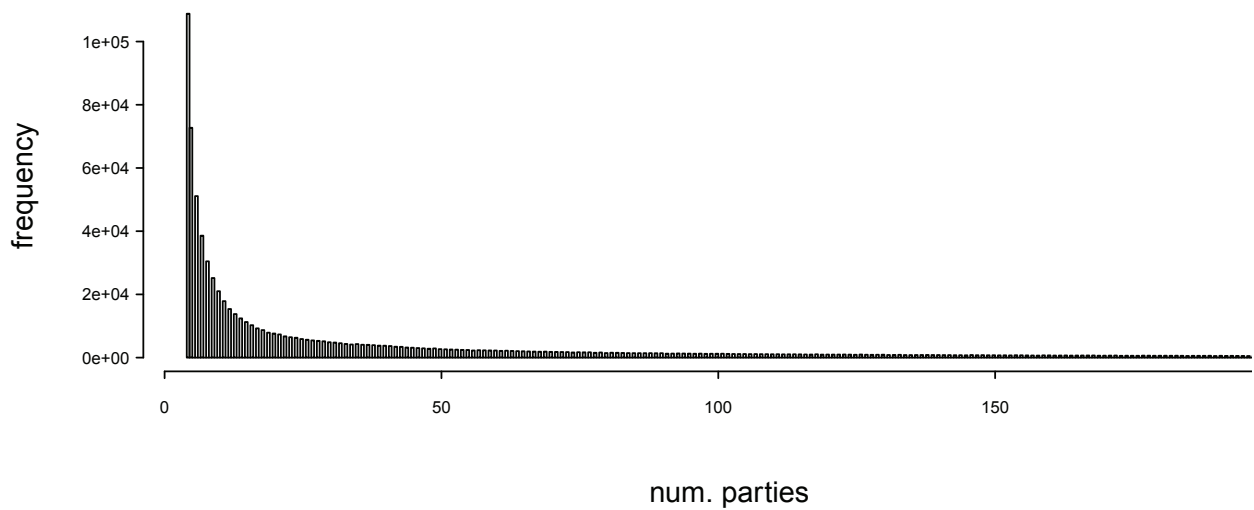}
\caption{Distribution of size for extracted communities.}
\label{FigSizeDistribution}
\end{figure*}

After removing duplicates, a total of $758,271$ communities were extracted from the network. Figure \ref{FigSizeDistribution} shows the distribution of size (number of parties) for extracted communities. Additional statistics are provided in Table \ref{TableCommunityStatistics}. As shown in Figure \ref{FigSizeDistribution}, the vast majority of extracted communities have more than 5 and less than 50 parties.

\begin{table*}[b]
\centering
\caption{Community statistics.}
\begin{tabular}{p{6cm} l}
\hline
total positive & $69,328$ (all positive)\\
total negative & $688,943$ (random sample)\\
mean num. parties & $33$\\
mean num. supplementary relationships & $47$\\
mean num. transactions & $124$\\
mean diameter & $3.5$\\
\hline
\end{tabular}	
\label{TableCommunityStatistics}
\end{table*}

\subsection{Results of Supervised Learning}

\begin{figure}
\centering
\includegraphics[width=0.45\textwidth,trim={0cm 0cm 0cm 2cm},clip]{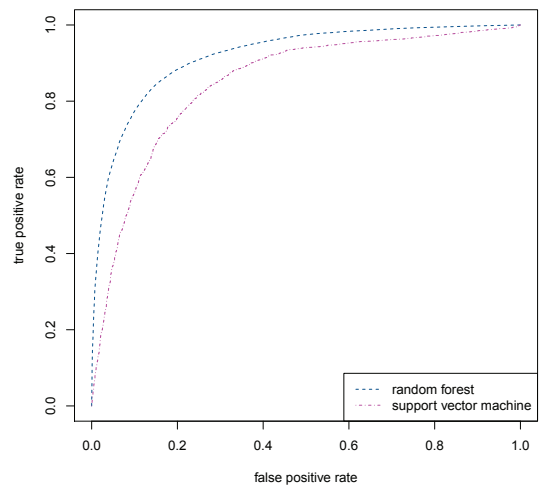}
\caption{ROC curve for random forest and SVM classifiers.}
\label{FigResultsRF}
\end{figure}

Table \ref{TableRFPerformance} gives the mean performance for a set of classifiers evaluated using the sampling process described in Section \ref{SectionMachineLearning}, taking $k = 10$. Example ROC curves are shown for a classifier of each model type. The results shown in Table \ref{TableNetworkSummary} and Figure \ref{FigResultsRF} indicate that both the random forest and SVM classifiers are able to achieve a level of performance that is suitable for use in a live environment. The random forest gives slightly better performance.

While the average recall of the classifiers is quite low, both models exhibit an extremely high precision. This means that a high classification threshold ($\tau$) can be selected, so that the classifier can operate with low rates of false positives. This is an important characteristic for real-world application of our system, as any communities classified as suspicious will be further investigated by human analysts. Since this is a time-consuming task, only minimal levels of false-negatives can be tolerated.



\begin{table*}
\centering
\caption{Mean performance of random forest classifiers over ten samples. Each random forest consisted of  100 trees. Parameters $\beta$ and $\tau$ refer to the weighting used in the calculation of F-scores and threshold used for classification, respectively.}
\begin{tabular}{p{3cm} p{1cm} p{1cm} p{1cm} p{1.5cm} p{1.5cm} l}
model & AUC & $\beta$ & $\tau$ & F-score & recall & precision\\
\hline
random forest & 0.92 & 0.1 & 0.93 & 0.96 & 0.31 & 0.98\\
& & 0.5 & 0.68 & 0.86 & 0.73 & 0.90\\
& & 1.0 & 0.47 & 0.85 & 0.88 & 0.82\\
SVM & 0.86 & 0.1 & 0.89 & 0.90 & 0.22 & 0.93\\
& & 0.5 & 0.63 & 0.80 & 0.70 & 0.83\\
& & 1.0 & 0.32 & 0.80 & 0.87 & 0.74\\
\hline
\end{tabular}
\label{TableRFPerformance}
\end{table*}

\section{Future Work}

The system described in this paper provides a basic framework for identifying money laundering activity in a transaction network. It is intended that future work will extend and improve this basic framework, informed by feedback provided from its use in an actual financial intelligence environment.

One limitation of our system is that network structure is represented solely through graph invariants. Moreover, the particular invariants used can only describe the global structure of each community. It may be that more localised descriptors, such as the role assignment proposed in \cite{Drezewski:2015fx}, provide a more informative view. In addition, our system considers the dynamics of each community only through analysis of transaction time-series. Future work will also consider the evolution of the community structure over time, and will attempt to capture relationships between the structure and particular edge and and vertex attributes.

Another limitation of our system is the reliance on expert knowledge for setting parameter values (e.g. thresholds imposed during community extraction). Future work will consider the sensitivity of our system to these values, and the use of automated approaches for determining optimal values.

Since our system uses supervised learning, it is unable to discover new typologies. To address this issue, future work will also consider unsupervised approaches. In particular, the use of network-based anomaly detection will be considered (see \cite{Savage:2014cc,Akoglu:2015kz} for recent reviews).


\section{Conclusions}

We have described an automated system for detecting money laundering operations in transaction networks.  This system advances the current state-of-the-art by analysing both explicit transaction relationships and implicit relationships derived from from supplementary information. The system extracts small, meaningful communities from this network in manner that allows existing business knowledge to be considered in the process. Supervised learning is then applied to these communities to obtain trained classifiers. Evaluation of the system shows that a suitable level of accuracy is achieved at high levels of precision. This is an important characteristic for our system, as use in a live environment necessitates a low rate of false positives.

\section*{Acknowledgments}

The authors wish to thank Mr Claude Colasante (AUSTRAC) for his tremendous support. This research is supported in part by an Australian Research Council Linkage Project in association with the Australian Transaction Reports and Analysis Centre (LP120200128).


\bibliography{references.bib}

\end{document}